\documentclass[a4paper,12pt]{article}
\usepackage{amsmath,amsfonts,amssymb,cite}

\begin{document}

{\small

\begin{center}
{\Large\bf A five-dimensional effective model for excited light mesons}
\end{center}

\begin{center}
{\large S.S. Afonin}
\end{center}

\begin{center}
{\it V. A. Fock Department of Theoretical Physics, Saint-Petersburg State
University, Russia}
\end{center}

\begin{abstract}
In the usual holographic approach to QCD, the meson spectrum is
generated due to a non-trivial 5-dimensional background. We propose
an alternative 5-dimensional scenario in which the spectrum emerges
due to coupling to a scalar field whose condensation is supposed to
be dual to the formation of gluon condensate and mimics the scale
anomaly in QCD. The spectrum of model has finite number of discrete
states plus continuum and reveals a Regge-like behavior in the
strong coupling regime.
\end{abstract}

\section{Introduction}

In the given report, we will discuss an effective five-dimensional
model for excited light mesons (basing on our paper~\cite{eff})
which represents an alternative to the standard bottom-up
holographic approach to QCD~\cite{AdS}. The main difference consists
in the mechanism for mass generation: Instead of mimicking the
spectrum as Kaluza-Klein excitations of 5D fields living in a
non-trivial 5D background we consider 5D fields in flat space
coupled to a scalar field which acquires non-zero vacuum expectation
value (v.e.v.) dependent on the fifth coordinate that has a physical
meaning of inverse energy scale. The scalar field is supposed to be
dual to the gluonic field strength tensor square. Such a
construction mimics the scale anomaly in QCD and generates a
discrete hadron spectrum with finite number of states plus
continuum.

We will regard our model as an effective field theory that is valid
below some energy scale, say below 2.5 GeV where the light mesons
have been detected~\cite{pdg}. Staying within this viewpoint, we do not need
the AdS metric in the UV limit as we do not perform matching to the
UV asymptotics of QCD correlators and the simplest possibility will
be considered --- the flat metric. In addition, as is the case of
many effective theories, our model is supposed to be applicable only
on the tree level even in the strong coupling regime.

\section{Vacuum sector}

Consider a scalar field $\varphi(x_{\mu},z)$, $\mu=0,1,2,3$, living
in five dimensions, where the 5th space-like coordinate $z$ is the
inverse energy scale, $0\leq z<\infty$. Let us assume that this field is dual
to the gluonic field strength tensor square $G^2_{\mu\nu}$. The latter is known to acquire a
non-zero v.e.v. $\langle G^2_{\mu\nu}\rangle$ which
causes the anomaly in the trace of energy-momentum tensor~\cite{anom} in QCD,
\begin{equation}
\label{anomaly}
4\varepsilon_{\text{vac}}=\langle \Theta_{\mu}^{\mu}\rangle_{\text{n.p.}}=
\frac{\beta(\alpha_s)}{4\alpha_s}\langle G^2_{\mu\nu}\rangle_{\text{n.p.}}+
\mathcal{O}(\alpha)+\cdots.
\end{equation}
where $\varepsilon_{\text{vac}}$ denotes the vacuum energy, n.p. is "nonperturbative part" and
the term $\mathcal{O}(\alpha)$ is the contribution of quark polarization effects.
We suppose that hadrons acquire
observable masses due to interaction with the field $\varphi$.
Taking the metric
$\eta_{AB}=(1,-1,-1,-1,-1)$, the action describing the pure vacuum
sector is postulated to be $(A=0,1,2,3,4)$
\begin{equation}
\label{1}
S_{\text{CSB}}=\int d^4xdz\left(\frac12\partial_A\varphi\partial^A\varphi+
\frac12m^2\varphi^2-\frac14\lambda\varphi^4\right).
\end{equation}
By assumption, the self-interaction is weak, $\lambda m\ll 1$,
hence, the semiclassical analysis may be applied.

The classical equation of motion is
\begin{equation}
\label{4}
\partial_{\mu}^2\varphi-\partial_z^2\varphi-\varphi(1-\varphi^2)=0.
\end{equation}
We also assume that the v.e.v. $\varphi_\text{vac}$ does not depend
on the usual space-time coordinates,
$\varphi(x_{\mu},z)=\varphi(z)$. The equation~\eqref{4} has then a
kink solution
\begin{equation}
\label{5}
\varphi_{\text{vac}}=\pm\tanh(z/\sqrt{2}).
\end{equation}
We choose the plus sign. The solution~\eqref{5} breaks
the translational invariance along the $z$-direction making thereby different energy scales
non-equivalent. The given effect is important at large enough $z$, {\it i.e.} at low energies,
at high energies the effect becomes negligible. This construction mimics the scale anomaly in
QCD, with the dependence on the fifth coordinate in the solution $\varphi_{\text{vac}}$
being related to the existence of anomalous dimension by the operator $G_{\mu\nu}^2$.

Consider the particle-like excitations of the vacuum state. Substituting
$\varphi=\varphi+\varepsilon$ into Eq.~\eqref{4}, retaining only linear in $\varepsilon$
contributions and assuming $\varepsilon(x_{\mu},z)=e^{ipx}\varepsilon(z)$, where $p^2=M^2$
is the usual 4D momentum defining the physical mass $M$, we obtain the following equation
for the discrete mass spectrum,
\begin{equation}
\left(-\partial_z^2+3\tanh^2(z/\sqrt{2})-1\right)\varepsilon_n=M_n^2\varepsilon_n.
\end{equation}
This one-dimensional Schr\"{o}dinger equation can be easily solved analytically, it is known
to have two normalizable discrete states (we omit the normalization factors),
\begin{equation}
\varepsilon_0=\frac{1}{\cosh^2(z/\sqrt{2})},\qquad M_0^2=0;
\end{equation}
\begin{equation}
\varepsilon_1=\frac{\tanh(z/\sqrt{2})}{\cosh(z/\sqrt{2})},\qquad M_1^2=\frac32.
\end{equation}
The continuum begins at $p^2=2$. The zero-frequency mode $\varepsilon_0$ corresponds to
the Goldstone boson of spontaneously broken translational invariance along the $z$-direction.
This "dilaton" mode is located at high energies, $z\rightarrow0$, where the scale invariance of the vacuum
is restored asymptotically. An interesting feature of the model is the existence of a massive mode
that is located near $z_0=\sqrt{2}\,\text{arctanh}(1/\sqrt{2})\approx1.25$ and could
be associated with a glueball according to the physical meaning of the field $\varphi$.

\section{Adding bosons}

Let us embed a massless hadron $H$ in the vacuum. The interaction
with the field $\varphi$ should generate a certain mass for the
hadron. We consider the simplest situation: The relevant action is
quadratic in $H$ and there is the factorization
$H(x_{\mu},z)=H(x_{\mu})H(z)$ that permits to obtain the
particle-like excitations. Since the field $H$ must be normalizable,
we can then easily integrate over $z$ and arrive at a 4D effective
action. We will not introduce the couplings of bosons to the
gradients of the scalar field $\varphi$ since such vertices,
although could play a role at high energies, are not important for
the mass generation at low energies.

For simplicity, we analyse a scalar field $\Phi$ coupled to the
vacuum field $\varphi$, the extension of the analysis below to the
higher spin fields is straightforward. The action is
\begin{equation}
\label{9}
S_{\text{bos}}=\int d^4xdz\left(\frac12\partial_A\Phi\partial^A\Phi-
\frac{G}{2}\varphi^2\Phi^2\right).
\end{equation}
Consider the particle-like
excitations $\Phi(x_{\mu},z)=e^{ipx}f(z)$, $p^2=M^2$ over the background~\eqref{5}.
The classical equation of motion represents a one-dimensional Schr\"{o}dinger
equation
\begin{equation}
\label{11}
\left(-\partial_z^2+\frac{G}{\lambda}\tanh^2(z/\sqrt{2})\right)f_n=M_n^2f_n,
\end{equation}
which determines the discrete spectrum
\begin{equation}
\label{12}
M_n^2=\frac12\left[\sqrt{1+\frac{8G}{\lambda}}\left(n+\frac12\right)-\left(n+\frac12\right)^2-\frac14\right],
\end{equation}
\begin{equation}
f_n=\cosh^{n-s}(z/\sqrt{2})
F\left[-n,2s+1-n,s+1-n,\frac{1-\tanh(z/\sqrt{2})}{2}\right]
\end{equation}
where $F$ is the hypergeometric function and
\begin{equation}
s=\frac12\left(\sqrt{1+\frac{8G}{\lambda}}-1\right),\qquad
n=0,1,2,\dots,\qquad n<s.
\end{equation}
The continuum spectrum sets in at $n=s$. The states with $n>0$ carry the same
quantum numbers as the ground state, in the language of potential models
they are referred to as radial excitations. These excitations emerge if
$G/\lambda>1$, the number of the radial excitations is controlled by the value
of coupling $G/\lambda$ to the field $\varphi$. An interesting feature of obtained spectrum
is that it is Regge-like, $M^2_n\sim n$, in the strong coupling regime $G/\lambda\gg1$.
A numerical fit for the vector case (the experimental situation with the scalar mesons
is ambiguous) is presented in Ref.~\cite{eff}

\section{Adding fermions}

Consider massless fermions coupled to the scalar field $\varphi$. Following
the standard procedure for introduction of fermions in 5D space,
the simplest action is
\begin{equation}
S_{\text{ferm}}=\int d^4xdz\left(i\bar{\Psi}\Gamma^A\partial_A\Psi-h\varphi\bar{\Psi}\Psi\right),
\end{equation}
where $\psi$ is a four component spinor and $\Gamma^{\mu}=\gamma^{\mu}$, $\Gamma^4=-i\gamma^5$,
here $\gamma^{\mu}$, $\gamma^5$ represent the usual 4D Dirac matrices.
Let us find the particle-like excitations over the
background~\eqref{5}. With the factorization $\Psi_{L,R}(x_{\mu},z)=e^{ipx}U_{L,R}(z)$ for the left
and right components, $\gamma_5\Psi_{L,R}=\pm\Psi_{L,R}$, the relation for masses follows from
the classical equation of motion,
\begin{equation}
\label{20}
\left(\pm\partial_z+\frac{h}{\sqrt{\lambda}}\tanh(z/\sqrt{2})\right)U_{L,R}=MU_{L,R}.
\end{equation}
At $z\geq0$, the equation~\eqref{20} possesses a normalizable zero-mode
solution describing a massless left-handed fermion,
\begin{equation}
M=0,\qquad U_L=\cosh^{-\frac{\sqrt{2}\,h}{\sqrt{\lambda}}}(z/\sqrt{2}),\qquad U_R=0.
\end{equation}
This mode is localized near $z=0$. If we considered the region $z<0$ the situation would be opposite:
The normalizable zero-mode solution describes a massless right-handed fermion. This feature
is in agreement with the interpretation of negative-energy solutions of the Dirac
equation as antiparticles. On the other hand, there is an asymptotic solution
\begin{equation}
z\rightarrow\infty:\qquad M=\frac{h}{\sqrt{\lambda}}, \qquad U_{L,R}=C_{L,R}.
\end{equation}
Here $C_{L,R}$ are some constants.

The generation of mass for the fermion in question can be obtained directly by means of
integration over $z$. With the help of parametrization
\begin{equation}
\Psi(x_{\mu},z)= s(z)\psi(x_\mu),\qquad  \int_{0}^{\infty}s^2(z)dz=1,
\end{equation}
we can integrate over $z$ (the zero-mode will not contribute)
and arrive at an effective 4D Lagrangian
\begin{equation}
\mathcal{L}_{\text{ferm}}^{(4)}=\bar{\psi}(i\gamma^\mu\partial_\mu-\mathcal{M})\psi,
\end{equation}
with
\begin{equation}
\mathcal{M}=\frac{h}{\sqrt{\lambda}}\int_{0}^{\infty}s^2(z)\tanh(z/\sqrt{2})dz.
\end{equation}
We note that $\mathcal{M}<h/\sqrt{\lambda}$ and the principal contribution to the effective
mass $\mathcal{M}$ comes from integration over large enough $z$, {\it i.e.} over the low-energy
region.

Thus, the fermion described by Eq.~\eqref{20} is massless and left-handed, this fermion
is localized only at high energies, at low energies its wave function is suppressed exponentially;
this phenomenon could mimic the confinement. In the limit of
zero energy, $z\rightarrow\infty$, the massless fermion disappears and instead a massive
fermion emerges that resembles a kink propagating like a particle.
It is natural to associate this solution with a light quark in QCD which looks practically
massless and left-handed if we probe it at high energies and acquires a constituent mass at low energies.
After integration over the energy scales the zero-mode is lost and the model describes
a massive fermion.

\section{Discussions and outlook}

The presented phenomenological model can serve as a basis for
constructing various effective models for QCD. We have considered an
application of the model to the calculation of boson masses and a
coupling of fundamental quarks to a field mimicking the physical
vacuum. The next natural applications of
the proposed approach is the description of the baryon spectrum. A
somewhat related question to be answered is the computing dynamical
information encoded in the correlation functions and finding
restrictions from the QCD sum rules in the narrow-width
approximation (see, {\it e.g.},~\cite{we} for a review).

The model has a natural limitation --- it should be regarded as an
effective model valid below some scale. As long as the model has
only one massive excitation that might be associated with the scalar
glueball, this scale could be the mass of the second scalar
glueball. Since this mass is expected at approximately the same
scale as the onset of perturbative continuum in light meson
spectrum, about 2.5~GeV, the model is able to describe the whole discrete meson
spectrum. As is the case of all effective models of QCD, the
considered model has no well established relation to QCD itself, but
rather mimics some features of the physics of strong interactions at
low and intermediate energies.

The five-dimensional model of low-energy strong interactions studied
here shares some general features with the bottom-up AdS/QCD models
(see, {\it e.g.},~\cite{AdS} and references
in~\cite{afonin}). First of all, the 5th space dimension is
introduced that has the physical meaning of inverse energy scale.
The equations determining the mass spectrum of hadrons are reduced
to the one-dimensional equations of the Schr\"{o}dinger type. The
holographic models, however, suffer from a large arbitrariness in
the choice of background metric and boundary conditions on the
holographic fields. As a consequence, any {\it ad hoc} spectrum can
be obtained, in particular, the spectrum of the
kind~\eqref{12}~\cite{regge}. The presented approach is much less
arbitrary since the fifth coordinate is singled out dynamically, as
a result one has a dynamical violation of the scale invariance while
within the usual holographic models this is implemented "by
hands". The considered approach represents thus
a scheme for building the low-energy models of QCD in which the
dependence on energy scale is included via the extension of usual
space-time to the fifth "energetic" dimension, the latter can be
always integrated out but we loose then the dynamical content of the
theory. It turns out that the introduction of "energetic" space
permits to calculate the spectrum of 4D theory and perhaps another
static properties of hadrons in an essentially semiclassical way,
such a possibility was the starting hypothesis of the AdS/QCD
conjecture.

\section*{Acknowledgments}

The work is supported by the Alexander von Humboldt Foundation and
by RFBR, grant 09-02-00073-a and travel grant 10-02-09280-mob$\_\,$z.

}


\begin{thebibliography}{9}
\bibitem{eff} S.~S.~Afonin, \emph{Int. J. Mod. Phys. A}, to be published; arXiv:0908.0457 [hep-ph].
%%CITATION = ARXIV:0908.0457;%%
\vspace{-2.5mm}
\bibitem{AdS} J.~Erlich, E.~Katz, D.~T.~Son, and M.~A.~Stephanov,
\emph{Phys. Rev. Lett.} \textbf{95}, 261602 (2005);
G.~F.~de Teramond and S.~J.~Brodsky, \emph{Phys. Rev. Lett.}
\textbf{94}, 201601 (2005);
L.~Da Rold and A.~Pomarol, \emph{Nucl. Phys.} \textbf{B721}, 79 (2005);
A. Karch, E. Katz, D. T. Son, and M. A. Stephanov,
\emph{Phys. Rev.} \textbf{D74}, 015005 (2006).
\vspace{-2.5mm}
\bibitem{pdg} C. Amsler {\it et al.}, \emph{Phys. Lett.}  \textbf{B667}, 1 (2008).
\vspace{-2.5mm}
\bibitem{anom} J.S.~Collins, L.~Duncan ,and S.D.~Joglekar,
\emph{Phys. Rev.} \textbf{D16}, 438 (1977).
\vspace{-2.5mm}
\bibitem{we} S. S.~Afonin, A. A.~Andrianov, V. A.~Andrianov and D.~Espriu,
\emph{JHEP} \textbf{0404}, 039 (2004).
%%CITATION = JHEPA,0404,039;%%
\vspace{-2.5mm}
\bibitem{afonin} S.~S.~Afonin, arXiv:1001.3105 [hep-ph].
%%CITATION = ARXIV:1001.3105;%%
\vspace{-2.5mm}
\bibitem{regge} S. S. Afonin, \emph{Phys. Lett.} \textbf{B675}, 54 (2009).
%%CITATION = ARXIV:0903.0322;%%
\end{thebibliography}
\end{document}